\documentclass[12pt]{article}
\usepackage[english]{babel}
\usepackage[latin1]{inputenc}
\usepackage{times}
\usepackage[T1]{fontenc}
\usepackage{epsfig}
\usepackage{amsmath}
\usepackage{amssymb}

\begin{document}

\title{Total screening and  finite range forces from ultra-massive scalar fields }

\author{H. Arod\'z$^a$,    J. Karkowski$^a$  and  Z. \'Swierczy\'nski$^b$ \\$\;\;$ \\ \emph{\small $^a$ Institute of Physics,
Jagiellonian University, Cracow, Poland }\\ \emph{\small
$^b$Institute of Computer Science and Computer Methods,}\\
\emph{\small Pedagogical University, Cracow, Poland}}

\date{$\;$}

\maketitle

\begin{abstract}
Force between static point particles coupled to a classical ultra-massive  scalar field is calculated.  The field potential is  proportional to the modulus of the field.   It turns out that the force exactly vanishes when the distance  between the particles exceeds certain finite value.  Moreover, each  isolated particle is  surrounded by a compact cloud of the scalar field that completely screens  its scalar charge.    
\end{abstract}

\vspace*{2cm} \noindent PACS:  11.27.+d, 11.10.Lm, 03.50.Kk \\

\pagebreak

\section{ Introduction}

The paradigmatic classical forces  between two elementary static charges in three-dimensional space are the Coulomb and Yukawa force, depending on whether the pertinent mediating field is massless or massive.  In the case of free fields, or weak nonlinear fields for which  field equations can be linearized,  there is no other possibility.  Nonlinearity of  the field equation can result in a very different force, e.g., the constant force confining quark and anti-quark.  Below we  present another  unusual  force  --  it  exactly vanishes
when the distance between the charges exceeds certain finite value  dependent on their strengths.  The mediating field is a classical scalar field with a V-shaped potential,  namely the  signum-Gordon field.  The name reflects the fact that the non-linear term in the field equation  is given by the $\mbox{sign}\:\varphi$ function, see Eq.\ (15)   below.   

Scalar fields  have found many applications in physics.   They appear in condensed matter physics  as effective fields  (scalar order parameters) in, e.g.,  Ginzburg -- Landau  type models of superconductors or atomic Bose -- Einstein condensates.  Effective scalar fields appear also in particle physics, e.g., in the Skyrme model of baryons. The discovery of the Higgs particle means that scalar fields are indispensable  in fundamental theories of matter. There is  a chance that scalar fields play important role  in cosmology.  Several scalar fields  have been considered in connection with  such phenomena  as the inflation or the presence of dark matter , e.g.,    axions,  chameleons,  and  K-fields  \cite{1}.  For instance,   the recent papers \cite{2}  are devoted to the  so called  classical fifth force due to various cosmological scalar fields.

Furthermore,  interesting scalar fields appear  also in a less spectacular,  but physically very sound  subject  of continuous descriptions of  discrete, nonlinear systems in the long wave limit. Here the classic example is the sinus-Gordon field  associated with a  long chain of   harmonically coupled pendulums \cite{3}.  The signum-Gordon field  considered in the present paper
gives an effective, long-distance description of a  system of coupled pendulums impacting on a stiff rod,   and also of a system of coupled balls elastically bouncing from a floor in the constant gravitational  field \cite{4, 5}. 
Earlier it was considered as a model of pinning \cite{6},  and  as a solvable model of decay of  false vacuum   \cite{7}.   The signum-Gordon field  is the relativistic scalar field $\varphi$ with the  V-shaped field potential given by the  absolute value of the field, 
$U(\varphi)=  g |\varphi|$,  where $g>0$ is the self-coupling constant.   Rather  surprisingly,   
models with  V-shaped potentials  have turned out to be quite interesting. In particular,  they support non-radiating compact oscillons  \cite{8},  as well as compact Q-balls \cite{9}.  The compact Q-balls have been generalized to  boson stars \cite{10} by including the gravity. Yet another  model with a V-shaped  potential was considered  in connection with baby-skyrmions \cite{11}.  Generally speaking, we think that the signum-Gordon field represents a new  class of scalar fields, which one may call the ultra-massive ones.  The point is that in the three-dimensional space  the  massless or massive fields are characterized by, respectively,  power-like or exponential approach to the vacuum value of the field, while in the signum-Gordon case the vacuum value is reached exactly,  on a  finite distance \cite{4}.   Such behavior is  typical for all scalar fields with a V-shaped field potential.

 The force between two point charges 
 in the case  of ultra-massive  mediating field has not been investigated yet.    Our main finding is that  the force has a strictly finite range.  Moreover,  it turns out that a single static charge is  surrounded by  a  compact spherical cloud of the scalar field which exactly vanishes outside certain ball centered at the charge -- the presence of the charge can be    
felt only at  close enough  distances.  It turns out that these effects are present also in the one-dimensional case. This is surprising because  in the massive and massless cases the change of dimensionality of space  significantly  influences  the asymptotic behavior  of the field.

The plan of our paper is as follows.  In  Section 2  we recall  the method  of computing the forces based on the energy-momentum tensor.   Section 3 is devoted to  the proper field of a static point charge and to the force between two static charges in the three-dimensional space.  In  Section 4 we consider the one-dimensional case, where we can obtain  exact analytic formula  for the force. Section 
5 contains a  summary and  remarks.

\section{Preliminaries }

The model we consider consists of    the dynamical real scalar field     $\varphi$   interacting with two  point-like  classical particles at rest.  The Lagrangian  has the form 
\begin{equation} L = \frac{1}{2} \partial_{\mu} \varphi  \partial^{\mu} \varphi
 - U(\varphi)  + j \varphi, \end{equation} where 
\begin{equation}
j=\sum_{l=1}^{2} q_l \: \delta(\vec{x} - \vec{a}_l). 
\end{equation}
Here $q_l$ is the scalar charge of the $l$-th particle and $\vec{a}_l$  its position in the space.    The field potential has the form $U= g |\varphi|$ in the case of signum-Gordon model, and  $U= m^2 \varphi^2/2$  in the case of free Klein -- Gordon field.  We use the  $c = \hbar =1$ units.    Because  the particles are kept at rest  we omit kinetic terms for them.   The Euler -- Lagrange equation corresponding to (1) reads
\begin{equation}
\partial_{\mu} \partial^{\mu} \varphi + U'(\varphi) = j. 
\end{equation}

In order to  find the force acting on the static charge $q_1$  we use the well-known method \cite{12} based on  computation of the total  flux of momentum through a closed surface  enclosing this charge with the other charge  left outside.  In particular,  it can be the sphere centered at $\vec{a}_1$ with arbitrarily small radius  $\epsilon >0$.   This flux gives the rate of transfer of momentum to the charge. It is equal to the force,  in accordance with the Newton's formula  $\vec{F} = d \vec{p}/dt$.  The momentum  density and its flux are given by the energy-momentum tensor for our system, 
\begin{equation}
T_{\mu\nu} = \partial_{\mu} \varphi  \partial_{\nu} \varphi   - \eta_{\mu\nu} L,
\end{equation}
where $(\eta_{\mu\nu}) = \mbox{diag}(1, -1,-1,-1)$ is the Minkowski space-time metric.

The total momentum $\vec{P}$ of the field  is given by 
\begin{equation}
P^i = - \int d^3x \:T_{0i} = - \int d^3x \:\partial_0 \varphi \partial_i \varphi. 
\end{equation}
We will consider only static fields, hence  $P^i =0$. The presence of the static charges  breaks the translational symmetry. Therefore, instead of the continuity equation we  have 
\begin{equation}
 \partial^{\mu} T_{\mu\nu} = -  \varphi \partial_{\nu} j. 
\end{equation}
The total flux of the $i$-th  component of the momentum  across a surface $\Sigma$  surrounding the charge $q_1$  is given by  $\int_{\Sigma} d \Sigma^k T_{ki}$.  Apart from the location  of the charges the flux is conserved, $\partial_k T_{ki}=0$, 
hence the total flux does not depend on the choice of $\Sigma$ provided that in the space between two such surfaces there are no charges.   The force exerted on the charge $q_1$ is equal to the total momentum flowing through the surface  $\Sigma$  enclosing this charge, 
\begin{equation}
F^i = - \int_{\Sigma} d \Sigma^k  \: T_{ki}, 
\end{equation}
where  the surface element  $d \vec{\Sigma} $ is directed outside the volume  $V$ enclosed by $\Sigma$.    The incoming  momentum  is  absorbed   entirely by the charge  because the momentum density of the field  remains constant --  in the case of static fields   $T_{0i} =0$.  The main  problem with formula (7) is that  in order to make use of it we first have to  find solution of  the field equation (3).  This can be a nontrivial task if the equation is nonlinear.  

Equations  (6), (7)   give in the static case
\begin{equation}
F^i = - \int_{V} d^3x \: \varphi \partial_i j = \int_{V} d^3x  \:\partial_i\varphi   j,  
\end{equation}
where we have used the fact that the charge  density $j$ vanishes on $\Sigma$.  Note that  these formulas for the force are correct 
when the integrands are well-defined, and   this is   not the case for point charges.  

In order to calculate  the force we will use formula (7)  with a conveniently chosen surface $\Sigma$.  The term  $j \varphi$ present in the Lagrangian $L$  and entering  $T_{ik}$, see formula (4), will vanish  on $\Sigma$.  In this way  we  avoid  problems with the  Coulomb type singularity of $\varphi$ at  the location of the point charge. 

As an illustration  and a check, let us compute the force between the two static point charges in the case of free scalar field with the potential $U(\varphi) = m^2 \varphi^2/2$.   The charge  $q_1$ is located at the origin, while  $q_2$ at the point $\vec{a}= (0,0, a)$.  Here $a >0$ gives the distance  between the charges.  Equation (3) reads 
\[ 
\Delta \varphi  - m^2  \varphi  = -  q_1 \delta(\vec{x}) -  q_2 \delta(\vec{x} - \vec{a}).
 \] 
Its vanishing at the spatial infinity solution has the form 
\begin{equation}
\varphi = q_1 \frac{e^{-m r}}{4 \pi r}  +  q_2 \frac{e^{-m |\vec{x} - \vec{a}|}}{4 \pi |\vec{x} - \vec{a}|}, 
\end{equation}
where $r = |\vec{x}|$.  In order to compute the force exerted on the charge $q_1$  we use formula (7)  with   $\Sigma$ the sphere of the small radius $\epsilon$,   $\epsilon \rightarrow 0$,  centered at the charge $q_1$.  The surface integral in (7) is simple in the spherical coordinates, especially in the limit  $\epsilon \rightarrow 0$. It gives  the following formula for the  force: 
\begin{equation}  
F^1=F^2=0, \;\;\; F^3= \frac{ q_1 q_2}{4 \pi} \left(\frac{m}{a} + \frac{1}{a^2}\right) \: e^{- m a}.
\end{equation}
We see that the dependence on the  distance $a$ has the  Yukawa form if $m^2 >0$,  and the Coulomb form if  $m^2=0$.  The charges of the same sign attract each other, and repel  if they have opposite signs.  The small distance behavior of the force  is obtained  by expanding with respect to the dimensionless product $a m$: 
\begin{equation}
 F^3=  \frac{q_1 q_2}{4 \pi a^2} \left[1 - \frac{(am)^2 }{2}  + \frac{(am)^3 }{3} - \frac{(am)^4 }{8}  + \dots \right].  
\end{equation}

In the case of static electric charges coupled to the electromagnetic field  the force can be calculated in the same manner. In particular, the electrostatic potential  $\psi$ also has the form (9),  with $m=0$  of course.  However, there is the sign difference 
between the symmetric stress tensors.  In the electrodynamics,  the  pertinent formula  has the form   \cite{13}
\[ T^{el}_{ik} =- E^i E^k - B^i B^k + \frac{1}{2} \delta_{ik}  ( \vec{E}^2 + \vec{B}^2).
\]
In the electrostatic case $E^k = - \partial_k \psi,  \: \vec{B}=0, $ and 
\[ T^{el}_{ik} =  - \partial_k\psi \partial_k \psi + \frac{1}{2} \delta_{ik}\: \partial_l \psi \partial_l \psi,
\]
while according to formula (4) for  the free massless scalar field   
\[ T_{ik} =   \partial_i \varphi \partial_k \varphi - \frac{1}{2} \delta_{ik}\: \partial_l \varphi \: \partial_l \varphi  
\]
(in the region of space in  which $j=0$).
Therefore, in the case of electric charges we obtain the expected repulsion of charges of same sign.

Let us digress on the case the mediating scalar field is tachyonic, i.e.,   $m^2 <0$.  In the presence of the static charges $q_1, q_2$  the field has the form 
\begin{equation} 
\varphi =  q_1 \frac{\cos(|m|r)}{ 4 \pi r}  +  q_2 \frac{\cos(|m| |\vec{x} - \vec{a}|)}{4 \pi |\vec{x} - \vec{a}|}, 
\end{equation}
which can be obtained from formula (9)  by inserting $m = i |m|$ and taking the real part.  The  force  turns out to be equal to the real part of  formula  (10), i.e., 
\begin{equation} 
F^1=F^2 =0, \;\; F^3  = \frac{  q_1 q_2}{4 \pi}  \left[  \frac{|m| \sin(|m| a)}{a}  +   \frac{\cos(|m| a)}{a^2}\right]. 
\end{equation}
In this case the force decreases very slowly with the distance $a$, and also its sign is not constant.  One should  however  remember that the field configuration (12) is unstable   --   even small perturbations can grow  exponentially with time when $m$ is imaginary, hence  the significance of formula (13) is  limited. 
The unstable modes  have the length of the  wave vector $\vec{k}$  bounded  by $|m|$,  $|\vec{k}| < |m|$. The corresponding wave length  $\lambda= 2 \pi/|\vec{k}|$ is bounded from below,   $ \lambda > \lambda_0= 1/2 \pi |m|$. 
If the distance $a$ is much smaller than  $\lambda_0$ the unstable modes may be regarded as almost homogeneous background field  generating a force which is very weak in comparison with (13).  Thus, formula (13)  may be regarded as relevant  when $ a |m| \ll 1/ 2 \pi $.  Its expansion  with respect to $ a|m|$ has the form  
\begin{equation}  F^3  =  \frac{q_1 q_2}{4 \pi a^2} \left[1 + \frac{ (a|m|)^2}{2}   - \frac{(a|m|)^4 }{8} +  \ldots \right]. \end{equation}

\section{The force in the  signum-Gordon model}

Exact calculation of the force mediated by a  self-interacting scalar field  most often  is not possible  because the pertinent exact static solutions of Eq.\ (3) are not available.  In the next section we show that  in the one-dimensional signum-Gordon model such solutions can be constructed.  In the three-dimensional case the exact formula for the force is not known. Nevertheless, the very fact that the force  exactly vanishes  when the   distance between the charges is large enough  can be seen rather easily -- it turns out that the proper fields  of the particles have strictly finite range.  

In the case of a single point charge $q>0$ located at the origin the field equation has the form 
\begin{equation}
\Delta \varphi = g\: \mbox{sign}\: \varphi - q\: \delta(\vec{x}),  \end{equation}
where the  $ \mbox{sign} $ function has the values $+1$  or $-1$ for $\varphi \neq0$, and  $ \mbox{sign}\: 0 =0 $.  Let us use an analogy between this equation and the Poisson equation of electrostatics.  The $\delta$ term generates the Coulomb part of the field, $\varphi_c = q/4 \pi r$, while the  $g\:  \mbox{sign}\:\varphi $  
term is interpreted as  to an `electric' charge density which is constant in the whole region in which $\varphi$ has a constant sign.  Close to the charge  the Coulomb part  dominates, hence the $\varphi >0$.  Here the nonlinear term effectively acts as  a constant negative charge density $-g$. The vanishing  `electric'  field 
$-\nabla \varphi$  is  obtained  when that charge exactly compensates the  charge $q$.  Thus, assuming the spherical symmetry,   we have   the condition 
\begin{equation}
q = \frac{4 \pi}{3} g R^3(q), 
\end{equation}  
which determines  the radius $R(q)$  of the ball in which we  have a nontrivial  field $\phi$.  Outside that ball the  `electrostatic' potential  is  put to zero by adding  the appropriate constant, thus $\varphi(r)=0$ for $r \geq R(q)$.   

Of course, such a heuristic reasoning does not  eliminate the necessity of  finding the  relevant solution $\varphi$ of  Eq.\ (15) inside that ball,  and checking whether  it can be matched with the vacuum field  $\varphi=0$. 
Assuming the spherical symmetry,  $\varphi$ obeys the equation
\begin{equation}
\partial^2_r \varphi + \frac{2}{r} \partial_r \varphi = g\: \mbox{sign}\:\varphi - q \: \delta(\vec{x}).
\end{equation}
The positive solution  that corresponds to  the heuristic picture above has rather  elementary, easy to guess form
\begin{equation}
\varphi_c(r) =   \frac{g}{6}  r^2 + \frac{q}{ 4 \pi r} + c_0(q),
\end{equation}
where $c_0(q)$ is  constant with respect to $r$, the subscript $c$ stands for `cloud'.   This solution can be glued with the vacuum 
field $\varphi=0$.   The matching conditions,  imposed at  a certain radius $R$,  have the standard form 
$ \varphi_c (R) =0, \; \partial_r \varphi_c (R) =0$.  They give  \[R= R(q)=\left(\frac{3 q}{4 \pi g}\right)^{1/3}\!,  \;\;\; c_0(q) = - \frac{3 q}{8 \pi R(q)}. \]   
To summarize,  the proper field of the located at the origin point charge  $q$ has the form 
\begin{equation}
\varphi_q(r) = \left\{\begin{array}{cc} \varphi_c(r) & \;\;\;  r \leq R(q),  \\  0 &  \;\;\;  r > R(q).   \end{array}  \right. 
\end{equation} 
The approach to the vacuum field when $r \rightarrow  R(q)$  from below  is quadratic: 
\[
\varphi_c(r) \cong \frac{g}{2} \:(R(q) -r)^2  \]
for $\; r \lesssim R(q)$.  This is typical for the V-shaped  self-interactions \cite{14}.

It is clear that in the case of  two point charges $q_1, q_2$  separated by a distance larger that $R(q_1) + R(q_2)$,   the pertinent solution of the field equation 
\begin{equation}
\Delta \:\varphi = g\: \mbox{sign}\: \varphi  - q_1 \delta(\vec{x}- \vec{b}) -  q_2 \delta(\vec{x} + \vec{b}),
\end{equation} 
where $2 |\vec{b}| > R(q_1) + R(q_2)$,  is simply the sum $\varphi_{q_1}(|\vec{r} -\vec{b}|)  +  \varphi_{q_2}(|\vec{r}+\vec{b}|)$.  The  charges  do not feel the presence of each other  at all  because there is  a region of pure vacuum  between them. One can surround each of the charges by a closed surface on which $T_{ik}=0$, hence the force  vanishes  exactly. 

When the distance between the charges is smaller than  $R(q_1) + R(q_2)$, the two clouds overlap and the nonlinearity of the $\mbox{sign}$  function becomes important.  For simplicity,  in the following we discuss  the symmetric  case $q_1 = q_2 \equiv q  >0$.  On the basis of  a hydrodynamical analogy presented below, we expect that the static solution of  Eq.\ (20)  represents a  compact static cloud of the  field   $\varphi >0$  which effectively screens the two charges.  Let us introduce $\vec{v} = - \nabla \varphi$ and write Eq.\ (20) in the form 
\[ \nabla \vec{v} = - g  + q \delta(\vec{x}- \vec{b}) +  q \delta(\vec{x} + \vec{b}) \]
(here $\mbox{sign}\:\varphi = +1$  because $\varphi >0$).  This equation can be regarded as  the governing equation for a stationary, free, irrotational  flow of a liquid with unit mass density and the local velocity $\vec{v}$.  The two $\delta$ terms  represent the point sources of the liquid, while the  constant $-g$ term  represents absorption of the liquid with the rate $-g$ per unit volume.  It is clear that  the flux rapidly decreases with the distance from the sources because of the absorption, and at a certain distance, which may depend on the direction in the space, it vanishes completely. Thus, there exists  a surface $\Sigma_{2q}$  enclosing the sources beyond which $\vec{v}=0$.  This means the  $\varphi $ has a constant  value  $\varphi_0 \geq 0$  on $\Sigma_{2q}$ and further away from the  sources.  Then, the function $\varphi - \varphi_0$  also obeys Eq.\ (20) inside $\Sigma_{2q}$, and it vanishes on 
$\Sigma$ together with its gradient.  The whole cloud has the total volume  equal to $|V(2q)| = 2q/g$. This  follows from   Eq.\ (20) because
\[ \int_{V(2q)}\!\! d^3x \: \Delta \varphi = \oint_{\Sigma_{2q}} \!\!d\vec{S} \:\nabla \varphi =0, \;\;\;  \int_{V(2q)} \!\!d^3x \: \mbox{sign}\:\varphi =  |V(2q)| . \]
Note that   $|V(2q)|$ does not depend on  $|\vec{b}|$.

Such a heuristic picture is quite convincing, nevertheless finding  the  exact shape of $\Sigma_{2q}$, and  the form of $\varphi$ inside the cloud is a different matter.  We are able to obtain approximate formulas  for the case the charges are close to each other, i.e., $|\vec{b}| \ll R(q)$.  When the two charges coincide ($\vec{b}=0$) the field has the form  (19) with $q$ replaced by $2q$, and the scalar cloud has the form of a ball. For  $\vec{b} \neq 0$ we expect an axially symmetric shape.  The general form of axially symmetric $\varphi$
obtained from Eq.\ (20)  under the assumption $\varphi >0$ reads 
\begin{equation}
\varphi(\vec{x}) = \frac{g}{6} r^2 + \frac{q}{ 4 \pi |\vec{x} - \vec{b}|}  +  \frac{q}{ 4 \pi |\vec{x} + \vec{b}|}  +  c_0(2q) + \delta c_0  +\sum^{\infty}_{l=1} \delta c_{2l} \; r^{2l}\: P_{2l}(\cos\theta),
\end{equation}
where the polar angle $\theta$ is counted from the $z$-axis that is chosen to be the straight-line on which lie the two charges.  It is convenient to put the origin of  the coordinate system ($r=0$)  half-way in between the charges, hence  $\vec{b} = (0,0,b),$  where $b>0$.   The  Legendre polynomials with odd indices do not appear in  (21)  because of the axial and reflection ($z \rightarrow -z$) symmetries of our set of charges.   The  unknown coefficients  $ \delta c_0, \delta c_{2l}$ in general depend on $b$, and they vanish for $b=0$.  The sum of the Coulomb terms is  even function of $b$, therefore we expect that  the perturbative coefficients  also are even functions of $b$, \begin{equation} \delta c_0 = b^2 \delta c_0^{(2)} + b^4 \delta c_0^{(4)} + \dots, \;\;\;\; \delta c_{2l} = b^2 \delta c_{2l}^{(2)} + b^4 \delta c_{2l}^{(4)} + \dots \end{equation}

Formula  (21)  holds inside the surface $\Sigma_{2q}$, which  in the spherical coordinates  is  conveniently  described  by the equation   
\[ r^2(\theta) = R^2(2q)\;(1+ \epsilon(\theta)), \] where  \[ \epsilon(\theta) = b^2 X^{(2)}(\theta) + b^4 X^{(4)} (\theta)  +  \dots \]  The coefficients (22) and the function  $\epsilon(\theta)$ are to be determined  from the conditions that ensure smooth matching of the solution (21) with the vacuum field $\varphi=0$  on $\Sigma_{2q}$:  
\[ \varphi(r(\theta)) =0, \;\;\; \left.  \partial_r \varphi\right|_{r(\theta)} =0. \]
These conditions are obtained  from Eq.\ (20) in the standard way, i.e.,  by  application of the Gauss theorem to the integral  of  $\triangle \varphi$ over  small  volumes  intersecting  with  the  surface  $\Sigma_{2q}$.  

Because  $b \ll r(\theta)$ in the region close to $\Sigma_{2q}$,  the Coulomb terms in (21) can be written as 
\[ \frac{q}{ 4 \pi |\vec{x} - \vec{b}|}  +  \frac{q}{ 4 \pi |\vec{x} + \vec{b}|} = \frac{q}{2\pi}   \sum^{\infty}_{l=0} \frac{ b^{2l}}{ r^{2l+1}}\: P_{2l}(\cos\theta). \]
After  simple calculations,  we obtain from the matching conditions considered to the order $b^2$ 
\begin{equation}
\delta c_0^{(2)}=0, \; \; \; \delta c_2^{(2)}=- \frac{q}{2 \pi R^5(2q)}, \; \;\; \delta c^{(2)}_{2l}=0 \;\; \mbox{for}\;\; l>1, \end{equation}
\begin{equation}
X^{(2)} = \frac{10}{3 R^2(2q)} P_2(\cos\theta).  \end{equation}
Calculations of the terms  of  the order $b^4$  and higher  are also rather straightforward,  but  we will not show the results  as they do not bring any really new features.   

The force exerted on the charge located at $\vec{x} = - \vec{b}$ can be obtained from formula (7) with  the plane $z=0$ as the surface $\Sigma$.  Because $\partial_z \varphi$ vanishes  on this plane,  formula (7) gives 
\[ F^1 =F^2=0,  \;\;\; F^3 =  \pi \int^{r(\frac{\pi}{2})}_0\! \!\! dr \; r \left[ (\partial_r \varphi)^2 + 2 g |\varphi| \right]. \]
It turns out that  with  the results (23), (24) one can compute $F^3$ up to the order $b^3$. The integrals are elementary, and we obtain
\[ F^3 = \frac{q^2}{16 \pi b^2} - \frac{g q }{3} b + \frac{2 g q}{3} \frac{b^3}{R^2(2q)} + \ldots . \]
Introducing the distance  between the charges   $a=2b$,  and inserting $R^{-2}(2q) = (2 \pi g/ 3q)^{2/3}$  we may also write 
\begin{equation}
F^3 = \frac{q^2}{4 \pi a^2} - \frac{g q }{6} a +  \left(\frac{\pi}{36 \sqrt 3}\right)^{\frac{2}{3}} \left( g^5  q\right)^{\frac{1}{3}} \; a^3 +  \ldots .
\end{equation} 
The first term  on the r.h.s. of this formula is the standard  attractive Coulomb force, present also in formulas (11), (14)  for the case of free  scalar field. It dominates at  small distances, i.e., when  $a \ll R(2q)$.  The other two terms do not have counterparts in formulas (11), (14).  The self-interaction of the scalar field results in the non-analytic dependence  both on the charge $q$ and the coupling constant $g$.

\section{The force in the one-dimensional signum-Gordon model}

The one-dimensional  case is interesting for two reasons. First,  we will see that the total screening effect persists, in spite of the  low dimension of the space.  This is rather surprising because  the Coulomb force in this case  does not vanish at all (it is constant),  and the Coulomb field of a single point charge linearly increases with the distance. 
Second,  we can obtain exact analytic formula  for the force, from which we see how it vanishes at a certain finite distance between the  charges.

Let us  first  have a look at  the field of a single point charge $q >0$  located at  $x=0$.  In one dimension the field equation 
has the form 
\begin{equation}
\partial^2_x \varphi  = g\: \mbox{sign}\:\varphi - q \: \delta(x).
\end{equation}
Integrating  both sides  of  this equation over infinitesimally small interval containing the point $x=0$  we obtain the condition
\begin{equation}
\lim_{\epsilon \rightarrow 0+} \left[  \partial_x \varphi(-\epsilon) -  \partial_x\varphi(\epsilon)\right] =  q. 
\end{equation}
Such discontinuity of the derivative $\partial_x \varphi$ is equivalent to the presence of point charge at 
$x=0$.  The pertinent  solution of Eq.\  (26)  is constructed piecewise,  by first solving the homogeneous equation 
\begin{equation}
\partial_x^2 \varphi = g \: \mbox{sign}\: \varphi
\end{equation}
in the two intervals $x <0$, $x >0$,   and next imposing the condition (27)  together with the continuity of $\varphi$ at $x=0$. 

 Boundary conditions at large $|x|$  have the form  $\varphi=0$  because
then the energy density, equal to $(\partial_x \varphi)^2/2 +g \: |\varphi|$, vanishes.   However there is a caveat to this  because the  $\mbox{sign}\:\varphi$ term  in  the  signum-Gordon equation (26)  does not approach  zero  unless  $\varphi=0$ exactly.    Therefore,  the asymptotic value $\varphi=0$  has to be reached exactly at a certain point at  finite distance from the charge.  It is  clear that the field  will stay equal to zero at larger distances  only if  the first derivative  $\partial_x\varphi$  also vanishes at the same point.

The  condition (27)  suggests that $\partial_x \varphi >0$ if  $x<0$, hence we expect    $ \varphi >0$  and  $\mbox{sign}\:\varphi = +1$  in a vicinity of the charge.   When $x\neq 0$,  Eq.\  (26)  is simplified to 
\begin{equation}
\partial_x^2 \varphi  =g,  
\end{equation} 
which has  the general solution of the form 
\begin{equation}     \varphi =  \frac{g}{2} x^2 + Ax +B,  \end{equation}
where $A, B$ are arbitrary constants.  This solution should match the trivial solution $\varphi=0$ at certain point $x=-d <0$. The matching conditions 
\[
\varphi(-d) =0, \;\; \left. \partial_x \varphi \right|_{x=-d} =0 \]
fix the constants $A,B$.  In this manner we obtain the solution in the interval  $-d \leq x < 0$:
\begin{equation}
\varphi_-(x)  = \frac{g}{2} \left(x+d\right)^2.
\end{equation}
Of course, $\varphi(x) =0$ for all $x \leq - d$.    In the region  $ 0 < x \leq d_1$   the solution has the form
\begin{equation}
\varphi_+(x)  = \frac{g}{2} \left(x-d_1\right)^2.
\end{equation}
This  function vanishes at $x=d_1$  together with its first derivative.   In the region  $x \geq d_1$ again  $\varphi=0$.  The continuity of $\varphi$  and the jump condition (27)  for $\partial_x \varphi$  at $x=0$
give 
\begin{equation} d_1=d= \frac{q}{2g}. \end{equation}
The full solution,  which may be called  
 the proper field of the charge $q$ and  denoted as $\varphi_q$, has the form 
\begin{equation} \varphi_q(x) = \left\{ \begin{array}{ll} 0 & \;\;\; x \leq -d,  \\  g (x+d)^2/2 & \;\;\; - d \leq x \leq 0, \\   g (x-d)^2/2 &  \;\;\; 0 \leq x \leq d,  \\ 0 & \;\;\; x \geq d.  \end{array}  \right. 
\end{equation}

In the case two point charges  $q_1$, $q_2$  located at the points $x=0$, $x=a$, respectively,  the field equation has the form 
\begin{equation}
\partial^2_x \varphi  =  g\: \mbox{sign}\: \varphi  - q_1 \delta(x) -  q_2 \delta(x-a).
\end{equation} 
Now we have two   conditions  of the form (27)
\begin{equation}
\lim_{\epsilon \rightarrow 0+} \left[ \partial_x \varphi(-\epsilon) -\partial_x\varphi(\epsilon)  \right] =  q_1, \; \lim_{\epsilon \rightarrow 0+} \left[ \partial_x \varphi(a-\epsilon) - \partial_x\varphi(a+\epsilon) \right] = q_2. 
\end{equation}
Similarly as in the case of single charge, we first solve the homogeneous equation  (28) 
in the three intervals: $x <0$, $\;0 < x  <a$, $\;a < x$,   next  we impose  the matching conditions  with the vacuum solution $\varphi=0$ at certain points $x=-d_-\leq 0$, $x=a + d_+$,  and finally we satisfy  the conditions (36)   together with the conditions of  continuity of $\varphi$ at $x=0$, $x=a$. 
The  form of the solution  crucially depends on the signs  and relative values of the charges  $q_1, q_2$.  
While it  is possible to obtain exact solution for arbitrary charges,  for the sake of simplicity we discuss here only the case $q_1=q_2 =q >0$, in which we expect  a non-negative  $\varphi(x)$ for all $x$.

Calculations  analogous to the ones carried out for the single positive charge $q$  give
\[
\varphi_-(x) = \frac{g}{2} (x + d_-)^2
\] in the interval $-d_- < x < 0$, where $d_- >0$. This solution matches the vacuum solution at $x=-d_-$, and  we have $\varphi(x)=0$ for all $x \leq -d_-$.   Similarly, in the interval  $a <x < a+d_+$  we have 
\[  \varphi_+(x) = \frac{g}{2} (x -a -  d_+)^2. \]
This solution merges with the vacuum solution at $x= a + d_+$.

It remains to determine $\varphi(x)$ in the interval  $0<x <a$.  We start from the general solution  (30). 
The  conditions at $x=0, x=a$ give four equations 
\[ g d_-^2 =2B, \;\; g d_- - A =q, \;\; g d_+^2 = g a^2 + 2 A a + 2 B, \;\; ga + A + g d_+ =q, \]
which determine the constants:
\[
d_+ = d_- = \frac{q}{g} - \frac{a}{2}, \;\;\;\; A = - \frac{ga}{2}, \;\;\;\; B= \frac{g}{2} \left(\frac{q}{g} - \frac{a}{2} \right)^2. \]  
Thus, 
\begin{equation}  \varphi_0(x) = \frac{g}{2} \left[\left(x-\frac{a}{2}\right)^2 + \frac{q}{g} \left(\frac{q}{g} -a \right) \right].
\end{equation}
Notice that also in this one dimensional case case the total effective charge of the scalar cloud, equal to $-g(d_- +a + d_+)$,  exactly compensates the total charge $2q$ of the point sources.  

It  remains to check  whether  $\varphi_0(x) >0$  for all $x \in (0,a)$,  as assumed when passing from Eq.\ (28) to  (29).   A glance at formula (37) reveals that this is the case only when the charges are not  too far from each other, namely when  
\[ a < a_*= \frac{q}{g}.\]
When the distance  $a$ exceed the critical value $a_*$  the cloud of the scalar field splits into two non-overlapping clouds of the form (34).  We se that $a_*= 2 d$, where $d$ is the half-width of  the cloud  for the single charge, formula (33).  

 To summarize,
in the case of two identical static point charges  $q>0$ located at the points $x=0$ and  $x=a >0$ the scalar field has the form 
\begin{equation} \varphi_{2q}(x) = \left\{ \begin{array}{ll} 0 & \;\;\; x \leq -d_-,  \\ \varphi_-(x) & \;\;\; - d_- \leq x \leq 0, \\ \varphi_0(x)  & \;\;\; 0 \leq x \leq a \\  \varphi_+(x) &  \;\;\; a \leq x \leq d_+ +a,  \\ 0 & \;\;\; x \geq d_+ + a.  \end{array}  \right. 
\end{equation}

In the one dimensional case   formula  (7) for the force  exerted on the charge  located  at $x=0$ has  the form 
\[ F^1 =  \left. T_{11}\right|_{x= -\epsilon} - \left.T_{11}\right|_{x=\epsilon},  \]  
where   $\epsilon$  can be any number from the interval $(0, a)$,    and $T_{11} = (\partial_x \varphi)^2/2 - g \: |\varphi|$  if $x \neq 0,  a$.  Using  the formulas for $\varphi_-, \: \varphi_0$   we  obtain  $  \left. T_{11}\right|_{x= -\epsilon} =0$, and a constant $  \left.T_{11}\right|_{x=\epsilon}$ (i.e., not dependent on $\epsilon$). Finally, 
\begin{equation}
F^1(a) = \frac{ q^2}{2}  \left(1- \frac{a}{a_{\star}}\right)
\end{equation}
for  $0<a \leq a_{\star}$ and  $F^1(a) =0$ if $a \geq a_{\star}$. 
Similar calculation yields the force exerted on  the charge $q$  located at  $x=a$. It is equal to  $-F^1(a)$, as expected.  The force  (39) is the attractive  one.  Of course, formula  (39)   does not apply to  the overlapping charges, i.e., when $a=0$.   In this case  the question about  exerted forces  becomes meaningless as the charges loose their identity.

\section{Summary and discussion}

1. We have found that  point charges  interacting with the signum-Gordon  classical scalar field are surrounded by a compact cloud of the field. This effect seems to  be independent of the dimension of the space. We have considered  the one- and three-dimensional case, but  generalization to other dimensions is straightforward. The compactness 
is typical for the ultra-massive fields,   e.g.,  the  Q-balls  and  the oscillons    in the signum-Gordon model  are compact too.  Outside the cloud the scalar field has exactly the vacuum value $\varphi =0$, as if the charge were absent.  The compactness of the cloud is directly related to the V shape of the potential  $U(\varphi) = g |\varphi|$ around  the vacuum field $\varphi=0$.  The ensuing $g \mbox{sign}\: \varphi$ term in the field equation  remains finite  even if the field is arbitrarily close to the vacuum field.  The only way to nullify it is to put $\varphi=0$ exactly.

When the clouds belonging to two charges  overlap, the  force appears. In the three-dimensional case  of two identical charges close to each other it is given by formula (25).  Comparing it with formulas  (11), (14)  for the force in the case of free Klein-Gordon  field we notice the peculiar  dependence on the strength $q$ of the charges.   This  is  due to the non-analytic dependence  on $q$ and on the coupling constant $g$  of the radius  $R(2q)= (3q/2 \pi g)^{1/3}$, as  demonstrated  by rewriting formula  (25)  in the form
\[F^3 = \frac{q^2}{4 \pi a^2} \left[1 -  \frac{a^3 }{R^3(2q)} +  \frac{1}{2} \frac{a^5 }{ R^5(2q)}+ {\cal O}(a^6)\right].  \]
In the one-dimensional case the force can be calculated exactly. It linearly decreases with the distance between charges until the critical distance $a_*$ is reached, at which the two clouds are completely  separated  and the force vanishes.

2. In the present work we have discussed the simplest case of  equal charges of same sign in order to focus on the main features of  the interaction with the ultra-massive scalar field.   In the one-dimensional case one can easily extend our analysis to general charges $q_1$, $q_2$.  Pertinent exact  formulas become much longer,  but no essentially new  features appear.  Of course, charges of the opposite sign repel each other.  In the three-dimensional case  the perturbative calculations are even more cumbersome.

3. The total screening of the charges by the compact clouds effectively makes the charged particles  neutral, i.e., their effective scalar charge vanishes, if measured for a distance.  In a sense,  the interaction mediated by our scalar field is of the contact type. The size of the scalar cloud $R(q)$ decreases with decreasing $q/g$,  hence it can be made very small.  Such a picture  can suggest ideas about dressing (some of)  the known particles in such clouds of the scalar field. We hope to explore such an exotic possibility in a future work.

\end{document}